# Molecular hydrogen beyond the optical edge of an isolated spiral galaxy

**Jonathan Braine & Fabrice Herpin**

*Observatoire de Bordeaux, UMR 5804, CNRS/INSU, B.P. 89, F-33270 Floirac, France*

Knowledge about the outermost portions of galaxies is limited owing to the small amount of light coming from them. It is known that in many cases atomic hydrogen (H I) extends well beyond the optical radius[1]. In the centres of galaxies, however, molecular hydrogen ($H_2$) usually dominates by a large factor[2–4], raising the question of whether $H_2$ is also abundant in the outer regions. Here we report the detection of emission from carbon monoxide (CO), the most abundant tracer of $H_2$, beyond the optical radius of the nearby galaxy NGC 4414. The host molecular clouds probably formed in the regions of relatively high H I column density and in the absence of spiral density waves. The relative strength of the lines from the two lowest rotational levels indicates that both the temperature and density of the $H_2$ are quite low compared to conditions closer to the centre. The inferred surface density of the molecular material continues the monotonic decrease from the inner regions. We conclude that although molecular clouds can form in the outer region of this galaxy, there is little mass associated with them.

The CO spectra shown in Fig. 1 reveal the presence of cool, but not very cold, molecular gas out to 1.5 times the optical radius ($R_{25}$) of the isolated spiral NGC 4414. The CO observations were carried out in good conditions with the 30-m antenna on Pico Veleta (Spain) operated by the Institut de Radioastronomie Millimétrique, in May and November 2003, and March 2004. Cool $H_2$ is not directly observable because it has no permitted rotational transitions so CO, the most abundant heteronuclear molecular, is used as a proxy. The galaxy chosen is expected to be representative of a large class of spirals because it is of late type, quite axisymmetric[5], and has no immediate neighbours with which it could interact. The atomic gas is extended, also indicating a lack of major interactions. This lack of interactions is important: although every interaction is different owing to the large number of parameters involved, isolated galaxies can reasonably be hoped to be representative. Furthermore, NGC 4414 is near the North Galactic Pole and suffers little from foreground (Galactic) confusion or extinction.

Almost no constraints are as yet available on the $H_2$ content of the outer disks of spirals. To our knowledge, these are the first published detections of CO beyond the optical radius of an external spiral, reaching $1.5 R_{25}$. A. Ferguson also reports detection (personal communication) of CO in the disk of NGC 6946 beyond $R_{25}$. Digel et al.[6] detected isolated clouds at large radii in the Milky Way, although the mass implied was quite low. Our previously published observations of NGC 4414, covering almost the whole optical disk, show a clear decline in the $CO(2 - 1)/CO(1 - 0)$ line ratio (independent of beam size) which we interpret[7] as a decrease in excitation temperature, meaning that even if substantial CO is present, the lines could be quite weak. Dust, which is necessary for $H_2$ formation, is clearly seen in absorption at the edge of the spiral disks of galaxies and even well beyond through the reddening of stars and background galaxies[8]. Observations of optical line ratios in H II regions at large radii also show that metals and star formation are present[9].

Figure 1 summarizes the positions for the positions (−8, 133), (−68, 162), (−98, 162), (31, −115), and (51, −138) in panels a to f. The main panel shows the galaxy NGC 4414 itself in R band with contours giving the H I column density. The circles indicate the beam size and the positions of the spectra shown individually. They are beyond the nominal optical radius $R_{25}$ and extend up to $1.5 R_{25}$ (−68, 162), with the position (−98, 162) not detected in CO.

The positions were chosen to maximize the likelihood of detecting CO by preferentially observing high H I column density regions. Indeed, the undetected position is quite close to a detected one but has only half the H I column density. This corroborates the findings of ref. 10, whose authors observed M33 with the BIMA interferometer in CO(1–0) and found molecular gas nearly exclusively in zones of high H I column density. They interpret this as showing that the $H_2$ forms from the high-density H I. M33 is a much smaller and lower metallicity spiral than NGC 4414, but there is also evidence for H I to $H_2$ conversion in the Milky Way[11,12]. A further indication that the $H_2$ is forming from H I in the outer parts of NGC 4414 is that although the CO line profiles are always narrower than the H I, the linewidths seem to be correlated (compare in Fig. 1). If the $H_2$ is indeed forming from H I in NGC 4414, then the $H_2$ formation process does not require a passage through spiral arms, because no spiral structure is observed in NGC 4414 (ref. 5).

Using a 'standard' $N(H_2)/I_{CO(1-0)} = 2 \times 10^{20}$ $H_2$ cm$^{-2}$ (K km s$^{-1}$)$^{-1}$ conversion ratio[13], the $H_2$ column density is 5–10% of the H I, roughly $N_{H\ I} \approx 5 \times 10^{20}$ cm$^{-2}$. The linewidths of the spectra are greater than those of giant molecular clouds and the molecular gas mass implied is (1–2) $\times 10^6 M_\odot$, depending on the position. In the outer parts the $N(H_2)/I_{CO(1-0)}$ ratio is probably somewhat higher[7,14] than the above-'standard' value which is used to enable comparison with previous work. It is commonly believed[15–17] that the H I is a two-phase medium, with the warm diffuse phase becoming more and more dominant at increasing galactocentric distances. Clearly, our results show that even where the stellar mass (light) contribution is very low (dim), $H_2$ formation is still possible. Whether this extends further out to lower, and possibly warmer, H I columns remains to be seen and is certainly necessary to determine the $H_2$ mass of spirals.

The CO detections presented here are really a continuous extension of the decline in CO brightness described in ref. 7, further confirmed by unpublished CO observations at radii slightly less than $R_{25}$. Analysis of the Hα image of ref. 18 shows that the Hα brightness also decreases continuously through the $R_{25}$ region, although the H I column density decreases rapidly[19]. It is generally believed—and confirmed for the Milky Way[14] and NGC 4414[7] through isotopic and dust continuum measurements—that the amount of $H_2$ is reasonably well known out to at least the solar circle or equivalent. Thus, for $H_2$ to represent as much or more mass than the H I at $R_{25}$ and beyond, the $H_2$ surface density must increase with radius over this range despite the regular decline in CO and Hα brightness and the (known) decrease in $H_2$ and H I surface densities at lower radii. From an analysis of the CO line data in these outer positions, we estimate the $H_2$ to represent at most 30% of the H I surface density at $1.5 R_{25}$; if $H_2$ were the dark matter in spiral galaxies[20], 50 times as much $H_2$ would be required to fit the rotation curve. □

# letters to nature

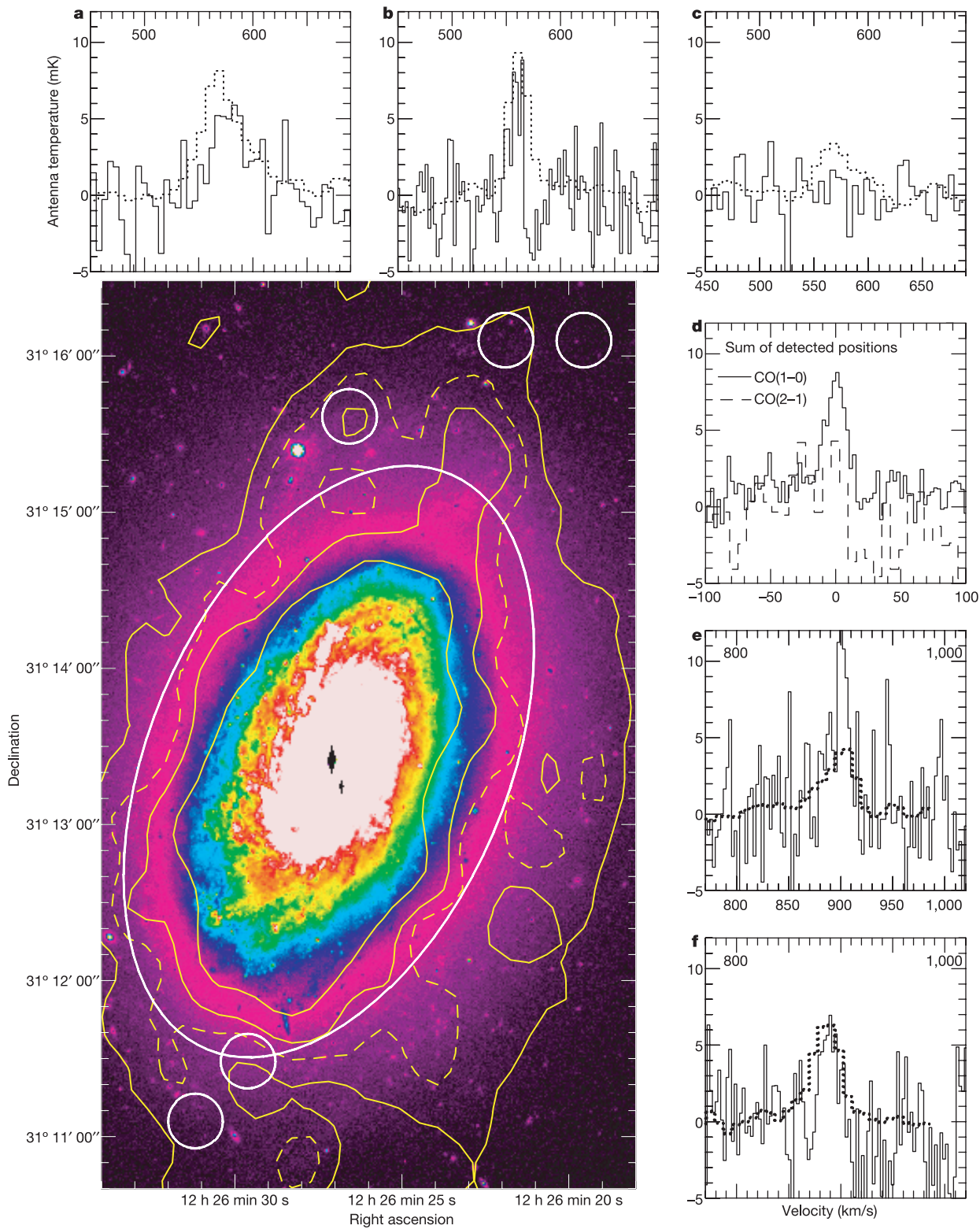

**Figure 1** Spectra showing the carbon monoxide and atomic hydrogen emission from the outer regions of NGC 4414. An R-band image of NGC 4414 taken with the CFHT is shown with H I contours from recent Westerbork observations at column densities of 4, 6, 8 and $12 \times 10^{20}$ atoms cm$^{-2}$, the $6 \times 10^{20}$ atoms cm$^{-2}$ contour being dashed. The ellipse indicates the average $R_{25}$ contour. The surrounding boxes show the CO (full line) and H I (dotted line) spectra for the positions indicated by the five circles. **a–c**, The positions (−8, 133), (−68, 162), and (−98, 162) are the top three circles. **e,f**, The positions (31, −115) and (51, −138) are the two lower circles. The CO(2−1) line was not clearly detected but the conditions were very good, enabling us to obtain a sensitive spectrum representing the sum of the four positions detected in CO(1−0). Panel **d** shows the CO(1−0) and CO(2−1) spectra summed over these positions. The line ratio is

CO(2−1)/CO(1−0) ≤ 0.5 (0.4 from the integrated intensities and 0.5 from fitting gaussians). The CO(1−0) line spectrum has been offset by +1 mK and the CO(2−1) by −1 mK for clarity. The beam widths differ by a factor of two for the two transitions but only if the CO is concentrated systematically either towards the centre of the beam (better picked up in CO(2−1)) or about 5–10″ away from the beam centre (that is, poorly picked up by the CO(2−1)) beam could the ratio be biased. Given that the linewidths are considerably greater than those of giant molecular clouds, the gas is not likely to be concentrated at any particular position with respect to the beam centre, rendering the ratio valid without further correction. All spectra are shown in the main beam temperature scale in millikelvins, the velocity being in kilometres per second (LSR, optical convention).

# letters to nature

**Acknowledgements** We thank J.-C. Cuillandre for taking the CFHT image and T. Osterloo, G. Gentile and G. Jozsa for making the H I data available. This work is based on observations carried out with the IRAM 30-m telescope. IRAM is supported by INSU/CNRS (France), MPG (Germany), and IGN (Spain).

**Competing interests statement** The authors declare that they have no competing financial interests.

**Correspondence** or requests for materials should be addressed to J.B. (braine@obs.u-bordeaux1.fr).